
\input harvmac
\def\half{{1 \over 2}}
\def\dz{{\partial_z}}

\def\dzb{{\partial _{\bar z}}}

\def \ad {{\dot a}}
\def \bd {{\dot b}}

\def \Gtp {{\tilde G^+}}
\def \Gtm {{\tilde G^-}}
\def \Gbp {{\bar G^+}}
\def \Gbm {{\bar G^-}}
\def \Gbtp {{\bar{\tilde G}^+}}
\def \Gbtm {{\bar{\tilde G}^-}}

\def \wGm {{\widehat {G^-}}}

\def \wGtp {{\widehat \Gtp}}

\def \wGbm {{\widehat {\Gbm}}}

\def \wGbtp {{\widehat \Gbtp}}

\def \pap {{\psi_\ad^+}}
\def \pam {{\psi_\ad^-}}
\def \pbp {{\psi_\bd^+}}
\def \pbm {{\psi_\bd^-}}
\def \pbap {{\bar\psi_\ad^+}}
\def \pbam {{\bar\psi_\ad^-}}
\def \pbbp {{\bar\psi_\bd^+}}
\def \pbbm {{\bar\psi_\bd^-}}
\def \euab {{\epsilon^{ab}}}

\def \euabd {{\epsilon^{\ad\bd}}}

\Title{\vbox{\hbox{IFUSP-P-1134}}}
{\vbox{\centerline{\bf Vanishing Theorems for the Self-Dual N=2 String}}}
\bigskip\centerline{Nathan Berkovits}
\bigskip\centerline{Dept. de Matem\'atica F\'isica, Univ. de S\~ao Paulo}
\centerline{CP 20516, S\~ao Paulo, SP 01498, BRASIL}
\centerline{and}
\centerline{IMECC, Univ. de Campinas}
\centerline{CP 1170, Campinas, SP 13100, BRASIL}
\bigskip\centerline{e-mail: nberkovi@snfma1.if.usp.br}
\vskip .2in
It is proven that up to possible surface terms, the only non-vanishing
momentum-dependent amplitudes for the self-dual N=2 string in $R^{2,2}$
are the tree-level two and three-point functions, and the only
non-vanishing momentum-independent amplitudes are the one-loop
partition function and the tree-level two and four-point functions.
The calculations are performed using the topological prescription developed
in an earlier paper with Vafa. As in supersymmetric non-renormalization
theorems, the vanishing proof is based on a relationship between the
zero-momentum dilaton and axion.

\Date{December 1994}
\newsec {Introduction}

In an earlier paper with Vafa\ref\bv{N. Berkovits and C. Vafa,
{\it N=4 Topological Strings}, preprint HUTP-94/A018 and
KCL-TH-94-12, to appear in Nucl. Phys. B, hep-th 9407190.},
a new topological prescription was
described for calculating scattering amplitudes of N=2 strings, and was
shown to be equivalent to the usual prescription.
Because the topological prescription does not require N=2
superconformal ghosts or integration over U(1) moduli, calculations are
considerably simplified. Furthermore, this new prescription contains no
ambiguities associated with the locations of the N=2 picture-changing
operators.

In the earlier paper \bv, it was proven for the self-dual string
in $R^{2,2}$ that all momentum-dependent amplitudes vanish up to surface
terms, with the exception of the three-point function.\foot{In an early
version of the preprint, it was also claimed that certain
momentum-independent amplitudes
vanish. However the proof of
this claim was incorrect since it ignored contractions between
$\partial_z x^\mu$ and $\partial_{\bar z} x^\mu$.
\ref\oo{H. Ooguri, private communication.}}
However there
exist indirect arguments that other amplitudes should also vanish.
Since a $g$-loop $N$-point amplitude can be cut into an $(N+2g)$-point
tree amplitude, one expects by ``unitarity'' arguments
\ref\dada{ A. D'Adda and F. Lizzi, Phys. Lett. B191
(1987) 85.}
that a loop amplitude
should vanish when the corresponding tree amplitude vanishes
(since there are two time directions, the
term ``unitarity'' should not be taken too literally). It was also argued
by Siegel\ref\sieg{W. Siegel, Phys. Rev. D47 (1993) 2504.}
that spacetime-supersymmetry and Lorentz-invariance imply
the vanishing af all loop amplitudes (note, however, that explicit calculations
find the one-loop partition function to be non-vanishing).

In this paper, it will be proven that up to possible surface terms, the
only non-vanishing momentum-dependent scattering amplitudes for the
self-dual string in
$R^{2,2}$ are the tree-level two and three-point functions, and
the only non-vanishing momentum-independent amplitudes are
the one-loop partition function and the tree-level
two and four-point functions.
This result appears to contradict an earlier one-loop calculation of the
three-point function\ref\ita{M. Bonini, E. Gava and R. Iengo,
Mod. Phys. Lett. A6 (1991) 795.}
and a two-loop calculation of the
partition function\ref\ovtwo{H. Ooguri and C. Vafa, private communication.}
which found non-vanishing amplitudes. A possible
resolution of this paradox is that these amplitudes
can be written as integrals of total derivatives. It would be
interesting to verify this fact with explicit calculations of the
surface-term contributions in the topological prescription. Note that these
calculations are
possible since the topological prescription does not contain
total-derivative ambiguities.

The vanishing proof in this paper will use the fact that inserting a
zero-momentum dilaton into a correlation function is related by
picture-changing to inserting a zero-momentum axion. It is interesting
that non-renormalization theorems for four-dimensional supersymmetric
strings also use the zero-momentum dilaton and axion, which are related
to each other by spacetime-supersymmetry transformations.\ref\sei{M. Dine
and N. Seiberg, Phys. Rev. Lett. 57 (1986) 2625.}
This suggests that the picture-changing operators, $\int G^+$ and
$\int \tilde G^+$, can be thought of as twisted spacetime-supersymmetry
generators, an idea that was first proposed in the earlier paper
with Vafa.\bv

For the self-dual string
\ref\adem{M. Ademollo, L. Brink, A. D'Adda, R. D'Auria,
E. Napolitano, S. Sciuto, E. DelGiudice, P. DiVecchia,
S. Ferrara, F. Gliozzi, R. Musto, R. Pettorino and J.H. Schwarz,
Nucl. Phys. B111 (1976) 77}
\dada
\ref\ov{H. Ooguri and C. Vafa, Nucl. Phys. B367 (1991) 83\semi
H. Ooguri and C. Vafa, Nucl. Phys. B361 (1991) 469\semi
H. Ooguri and C. Vafa, Mod. Phys. Lett. A5 (1990) 1389.}
in $R^{2,2}$, the worldsheet fields are
$X_{a \dot b}$, the right-moving $\psi_{\dot a}^+$ and $\psi_{\dot a}^-$,
and the left-moving $\bar\psi_{\dot a}^+$ and $\bar\psi_{\dot a}^-$.,
Note that the SO(2,2) vector index is expressed as two SU(1,1) spinor
indices, $a$ and $\dot a$, which take the values
$1$ or $2$. In this notation, $x_{a\ad}=x_\mu \sigma^\mu_{a\ad}$
where
$$\sigma^0_{a\ad}=i\delta_{a\ad},\quad
\sigma^1_{a\ad}=i\sigma^x_{a\ad},\quad
\sigma^2_{a\ad}=i\sigma^y_{a\ad},\quad
\sigma^3_{a\ad}=\sigma^z_{a\ad}.$$ SU(1,1) indices can be raised
and lowered using the epsilon tensor in two dimensions.

After twisting, the superconformal constraints are:
\eqn\const{L= {1\over 4}\dz x_{a\dot b}\dz x^{a\dot b}+\euabd \pam\dz\pbp,}
$$G^+= \pbp \dz x_{2}^\bd, \quad
\tilde G^+= \pbp \dz x_1^\bd, \quad
G^-= \pbm \dz x_1^\bd, \quad
\tilde G^-= \pbm \dz x_2^\bd, $$
$$J^{++}=\half\euabd\pap\pbp,\quad
J=\euabd\pap\pbm,\quad
J^{--}=\half\euabd\pam\pbm,$$
$$\bar L={1\over 4}\dzb x_{a\dot b}\dzb x^{a\dot b}+\euabd \pbam\dzb\pbbp,$$
$$\bar G^+= \pbbp \dzb x_2^\bd, \quad
\bar{\tilde G}^+= \pbbp \dzb x_1^\bd, \quad
\bar G^-= \pbbm \dzb x_1^\bd, \quad
\bar{\tilde G}^-= \pbbm \dzb x_2^\bd, $$
$$\bar J^{++}=\half\euabd\pbap\pbbp,\quad
\bar J=\euabd\pbap\pbbm,\quad
\bar J^{--}=\half\euabd\pbap\pbbm,$$
which form a right-moving and left-moving twisted N=4 superconformal algebra.

In the topological description of the N=2 string, the physical vertex
operators contain
U(1) charge $(1,1)$, are annihilated by $\int G^+$, $\int\tilde G^+$,
$\int \bar G^+$, $\int\bar{\tilde G}^+$, and can not be written as
$\int G^+ \tilde G^+ \Omega$ or $\int \bar G^+ \bar{\tilde G}^+\Omega$
for any $\Omega$. They consist of the momentum-dependent state
$$G^+ \Gbp V=k^{1\ad}k^{1\bd}\pap\pbbp e^{ik\cdot x}$$
and the momentum-independent
states $\pap\pbbp$. Note that since $k\cdot k=0$, $k^{2 \ad}$ is proportional
to $k^{1 \ad}$ so there is only one momentum-dependent physical state.

It will first be proven that
except for the tree-level two and three-point functions,
all momentum-dependent amplitudes must vanish up to surface terms.
It will then be proven that except for the one-loop partition function and
the tree-level two and four-point functions,
all momentum-independent amplitudes
must vanish up to surface terms.

\newsec {Vanishing Theorem for Momentum-Dependent Amplitudes}

When there are momentum-dependent vertex operators, it was shown in
reference \bv that the $g$-loop $N$-point
amplitude on a surface of instanton number
$(2g-2+N, 2-2g-N)$ can be expressed as:

\eqn\momdep{
A= \int_{M_{g,n}}\prod_{i=1}^{3g-3+N} G^-(\mu_i) \Gbm(\bar\mu_i)
\quad
\prod_{r=1}^{N-1} \phi_r (z_r)
\quad e^{ik_N\cdot x(z_N)}}
$$
\det(Im \tau)
 |\det \omega^k (v_l)|^2 \prod_{j=1}^g \Gtp(v_j)\Gbtp(\bar v_j) $$
where $\mu_i$ are the Beltrami differentials and
$\phi_r$ is a physical vertex operator of U(1) charge (1,1) of the
type described in the introduction.
Note that $\Gtp$ and $\Gbtp$ have no poles, so the
amplitude is independent of the locations of the $v_j$'s.

The only amplitude that can not be expressed in this form is the
tree-level two-point function since $3g-3+N<0$. In this case, the
amplitude is given by
$$<e^{ik_1\cdot x(z_1)} e^{ik_2\cdot x(z_2)}
J^{++}(z_3) \bar J^{++}(\bar z_3)>.$$

As was shown
in reference \bv,
amplitudes on surfaces of other instanton numbers $(I_R,I_L)$
are related
to $A$ by
\eqn\instanton{A_{I_R,I_L}= h^{I_R-I_L-4g+4-2N}A}
where $h k_N^{2\ad}=k_N^{1\ad}$ (note that $h \bar h=1$).
Therefore $A$=0 implies that $A_{I_R,I_L}=0$ for surfaces of all instanton
numbers.

The first step in proving that $A$ vanishes is to insert a zero-momentum
dilaton vertex operator into the amplitude. This vertex operator is given
by $\int_\Omega d^2 z (\dz x^\mu \dzb x_\mu)$ where the region of integration,
$\Omega$, covers the whole surface with the exception of small discs
surrounding $\mu_i$, $v_j$, and $z_r$. As was shown in appendix B of
reference
\ref\atick {J. Atick, G. Moore, and A. Sen, Nucl. Phys. B307 (1988) 221.},
this amplitude with the insertion is equal to
\eqn\dil {A^{dilaton}=(gd-\sum_{r=1}^N k_r \cdot {d\over d\hat k_r} )A}
where $g$ is the genus, $d$ is the dimension ($d=4$), and
$d/d\hat k^\mu$ acts only on the $k$'s appearing in the exponential
$e^{ik\cdot x}$ of the vertex operators, but not on the $k$'s appearing
as factors in front of the exponential (for example,
$d/d\hat k^\mu
(k^{1\ad} \psi^+_\ad)(k^{1\ad}\bar\psi^+_\ad) e^{ik\cdot x}=
i x_\mu
(k^{1\ad} \psi^+_\ad)(k^{1\ad}\bar\psi^+_\ad) e^{ik\cdot x}$).

As discussed in reference \atick, the term $gd$ in equation \dil
comes from the
contraction of $\dz x^\mu$ with $\dzb x^\mu$ in the dilaton vertex
operator. The term
$-\sum_{r=1}^N k_r\cdot d/d \hat k_r$
comes from writing the dilaton vertex operator as the surface term
$$\int d^2z \dz(x_\mu \dzb x^\mu)=\half \int_C d\bar z \dzb (x_\mu x^\mu)$$
where the contour $C$ surrounds the points $\mu_i$, $v_j$, and $z_r$.
Since the operators at $\mu_i$ and $v_j$ only involve derivatives of
$x^\mu$, their surface terms vanish. However at $z_r$, the surface
term contributes $-(ik_r\cdot x) \phi_r=-k_r\cdot d/d\hat k_r \phi_r$
where the $d/d\hat k$ acts only on the $k_r$ in the exponential.

Because the $\dz x_{1\ad}$'s and $\dzb x_{1\ad}$'s
in $G^-$, $\Gtp$, $\Gbtm$, and $\Gbp$ can not contract with each other,
they can only be contracted with the $x^\mu$'s appearing in the exponentials
of $\phi_r$, which bring down factors of $\hat k_r^\mu$.
Therefore $\sum_{r=1}^N
k_r\cdot d/d\hat k_r A= ((3g-3+N)+g+(3g-3+N)+g)A$. Since $d=4$, equation
\dil implies that
\eqn\prop{A^{dilaton}=(6-4g-2N)A.}

The next step in the proof is to show that $A^{dilaton}=0$. This is done
by writing the integrand of the dilaton vertex operator in the form
$$\dz x^\mu \dzb x_\mu={\euabd \over 2} (\dz x_{1\ad}
 \dzb x_{2\bd} -\dz x_{2\ad}\dzb x_{1\bd})$$
\eqn\vert{={\euabd \over 2} ((\Gtp\pam) \dzb x_{2\bd} -\dz x_{2\ad}
(\Gbtp \pbbm)).}

In the first term of equation
\vert, the $\Gtp$ can be pulled of the $\pam$ to encircle the
vertex operator
$V_N(z_N)=e^{ik_N\cdot x(z_N)}$, where possible surface terms are
being ignored. Since $(\Gtp V_N)= -h(G^+ V_N)$ where $k_N^{2\ad}
=h k_N^{1\ad}$,
the $G^+$ can be pulled back on the $\pam$ to give
\eqn\first{{\euabd\over 2} (-h G^+\pam) \dzb x_{2\bd} =
-h{\euabd\over 2} \dz x_{2\ad} \dzb x_{2\bd} .}

Similarly for the second term of equation \vert,
the $\Gbtp$ can be pulled onto
$V_N$, replaced with $-h \Gbp$, and returned to encircle $\pbbm$. This gives
$$-{\euabd\over 2} \dz x_{2\ad} (-h\Gbtp \pbbm)=
h{\euabd\over 2} \dz x_{2\ad} \dzb x_{2\bd} $$
which cancels the contribution in equation \first.

So we have proven that $A^{dilaton}$=0, which implies that either
$(6-4g-2N)=0$ or $A$=0. So besides the tree-level two-point function,
the only possible non-zero amplitudes (up to surface terms) are when
$N=1,g=1$ or $N=3,g=0$. But by momentum conservation, the one-point
amplitude can not contain momentum dependence. So up to surface terms,
the only non-vanishing momentum-dependent amplitudes for the
self-dual string are the tree-level two and three-point functions.

\newsec {Vanishing Theorem for Momentum-Independent Amplitudes}

It will now be proven that up to surface terms, all momentum-independent
amplitudes must vanish
except for the one-loop partition function and the tree-level
two and four-point functions. As was shown in reference \bv,
momentum-independent amplitudes can be written in the form
\eqn\momindep{
F(u_R,u_L)=\int_{M_{g,N}}\prod_{i=1}^{3g-4+N}\wGm (\mu_i)\wGbm(\bar\mu_i)
\quad J^{--}(\mu_{3g-3+N}) \bar J^{--}(\bar\mu_{3g-3+N})}
$$\prod_{r=1}^N \phi_r(z_r)\quad
\det(Im\tau)|\det \omega^k(v_l)|^2
\prod_{j=1}^{g}\wGtp(v_j)\wGbtp(\bar v_j)
$$
where
$$\wGm=u^R_a \dz x^{a\ad} \pam=u_1^R \Gtm -u_2^R G^-,\quad
\wGtp=u^R_a \dz x^{a\ad} \pap=u_1^R G^+ -u_2^R \Gtp,$$
$$\wGbm=u^L_a \dzb x^{a\ad} \pbam=u_1^L \Gbtm -u_2^L \Gbm,\quad
\wGtp=u^L_a \dzb x^{a\ad} \pbap=u_1^L \Gbp -u_2^L\Gbtp,$$
$u^R_a$ and $u^L_a$ are SU(1,1) spinors which parameterize the choice
of complex structure, $\phi_r$ are momentum-independent vertex operators
of the form $\pap\pbbp$, and $F(u_R,u_L)$ is a polynomial of
degree $(4g-4+N,4g-4+N)$ in $(u_R,u_L)$ whose $(8g-7+2N)^2$ components give the
scattering amplitude on a surface of instanton-number $(I_R,I_L)$
where $-4g-4+N\leq I_R,I_L\leq 4g-4+N$.

The only amplitudes that can not be expressed in this form are the
tree-level two-point function and the one-loop partition function. In
these cases, the amplitudes are defined as
$<\phi_1(z_1)\phi_2(z_2)>$ and as
$\int_{M_1}
(\int d^2 z J(z) \bar J(\bar z))^2$.

Since $F$ is invariant under the SU(1,1) subgroup of SO(2,2)
Lorentz transformations which
transform $u^R_a$ and $u^L_a$
but leave $\psi_\ad^\pm$ and $\bar\psi_\ad^\pm$
invariant, $F$ must be proportional
to $(\euab u^R_a u^L_b)^{4g-4+N}$. It will now be shown that the
proportionality constant vanishes up to surface terms unless $N=2-2g$
or $N=4-4g$.

The first step is to compute the effect of inserting the zero-momentum
axion vertex operator
\eqn\axion{V_{ab}=
b^{ab}\int_{\Omega} d^2 z (\dz x_{a\ad} \dzb x_b^\ad)}
where $b_{ab}=b_{ba}$ is the polarization of the axion, and
the region of integration, $\Omega$, covers the whole surface
with the exception of small discs surrounding $\mu_i$, $v_j$, and $z_r$.
One can write the vertex operator as the surface term
$$\half b^{ab}(\int_C d\bar z(x_{a\ad} \dzb x_b^\ad)
-\int_C dz (x_{a\ad}\dz x_b^\ad))$$
where the contour $C$ surrounds the points $\mu_i$, $v_i$, and $z_i$.

It is easy to check that this surface term transforms
$\dz x_{a\bd} \to 2 b_{ab}\dz x^b_\bd$ in $\wGm$ and $\wGtp$,
transforms
$\dzb x_{a\bd} \to -2 b_{ab}\dzb x^b_\bd$ in $\wGbm$ and $\wGbtp$, and
leaves $\phi_r$ invariant.
The amplitude with the axion insertion therefore satisfies the
following identity:
\eqn\ident{A^{axion}=2 b^{ab}\epsilon_{bc}
 (u_a^R {d\over du_c^R} -u_a^L{d\over du_c^L}) A.}

So the amplitude with an insertion of
$\int d^2 z \dz x_{1 \ad} \dzb x_1^\ad$ satisfies the identity
\eqn\Aone{A^{b_{11}}=2(u_1^R d/du_2^R -u_1^L d/du_2^L) A.}
Furthermore,
since
\eqn\sum{\int d^2 z \dz x_{2\ad}\dzb x_{1}^\ad=}
$$- \int d^2 z
\dz x^\mu \dzb x_\mu +\half
\int d^2 z (\dz x_{1\ad}\dzb x_{2}^\ad
+\dz x_{2\ad}\dzb x_{1}^\ad),$$
the amplitude with an insertion of
$\int d^2 z \dz x_{2\ad}\dzb x_{1}^\ad$ satisfies the
identity
\eqn\Atwo{A^{dilaton + b_{12}}=
-[gd+ (u_1^R d/du_1^R -u_2^R d/du_2^R -u_1^L d/du_1^L
+u_2^L d/du_2^L)]A}
where the term $gd$ comes from the dilaton contribution as was
explained in the previous section.

But
$\int d^2 z \dz x_{1 \ad} \dzb x_1^\ad$ is related
by picture-changing to an insertion of
$\int d^2 z \dz x_{2\ad}\dzb x_{1}^\ad$.
In other words,
$\int d^2 z \dz x_{1 \ad} \dzb x_1^\ad$=
$G^- W$ and
$\int d^2 z \dz x_{2\ad}\dzb x_{1}^\ad$=
$\tilde G^- W$ where
$W=\psi_\ad^+ \dzb x_1^\ad$. As was shown in reference \bv,
this implies that up to surface terms,
\eqn\harm{{d\over du_1^R}A^{b_{11}}=
-{d\over du_2^R}A^{dilaton +b_{21}}.}

Using equations \Aone, \Atwo, and \harm, and the fact that
$A$ is proportional to $(u_1^R u_2^L -u_2^R u_1^L)^{4g-4+N}$,
it is straightforward to show that
$(2g-2+N) (4g-4+N) A=0$. So the amplitude must vanish unless
$N=2-2g$ or $N=4-4g$.

Therefore up to surface terms, the only non-vanishing
momentum independent amplitudes are the one-loop partition function
and the tree-level
two and four-point functions.
It would of course be interesting to check if these results
are spoiled by surface-term contributions.

{\bf Acknowledgements:}
I would like to thank Denis Dalmazi, Hirosi Ooguri, Victor
Rivelles, and Cumrun Vafa for useful
conversations, and the Conselho Nacional de Pesquisa for
financial support.
\listrefs
\end